\documentstyle[rotate,times,pramana,floats,psfig]{ias}
\input {rotate.tex}
\input {isolatin1.sty}
\input {wasysym.sty}
\input {pifont.sty}

\begin{document}

\def\chargedensity{\rho}

\def\E{{\bf E}}

\def\Q{{\bf Q}}

\def\J{{\bf J}}

\def\B{{\bf B}}

\def\D{{\bf D}}

\def\r{{\bf r}}

 \def\v{{\bf v}}

 \def\dV{{\; \rm d^3}{\bf r}}

\def\curl{{{ \rm curl}\; }} 

\def\grad{{{ \rm grad}\; }} 

\def\div{{{\rm  div}\; }} 

\def\p{{\bf p}} 

\def\U{{ U}} 

\def\dv{{\rm  d^3}{\bf r}}

 \def\O{{ O}}
 
 \def\Z{{\mathcal Z}}
 
 \mark{{Local Simulation Algorithms}{Levrel et al.}}  \title{Local
   Simulation Algorithms for Coulombic Interactions} \author{ L.
   Levrel$^1$, F. Alet$^2$ J. Rottler$^3$, A.C. Maggs$^1$, } 
\address{
   $^1$ Laboratoire de Physico-Chimie Th\'eorique, UMR CNRS-ESPCI
   7083, 10 rue Vauquelin, F-75231 Paris Cedex 05, France.\\
   $^2$Theoretische Physik and Computational Laboratory, ETH Zürich,
   CH-8093 Zürich, Switzerland.\\
   $^3$ PRISM, Bowen Hall, Princeton University, Princeton, NJ 08544, USA.\\
 }
 
 \pacs{2.0} \abstract{ We consider  dynamically
   constrained Monte-Carlo dynamics and show that this leads to the
   generation of long ranged effective interactions.  This allows us
   to construct a local algorithm for the simulation of charged
   systems without ever having to evaluate pair potentials or solve
   the Poisson equation.  We discuss a simple implementation of a
   charged lattice gas as well as more elaborate off-lattice versions
   of the algorithm.  There are analogies between our formulation of
   electrostatics and the bosonic Hubbard model in the phase
   approximation.  Cluster methods developed for this model further
   improve the efficiency of the electrostatics algorithm.  }
 \maketitle
\section{Introduction}

Computer modeling of charged systems is demanding due to the range of
the Coulomb interaction\cite{schlick}. The direct evaluation of the
Coulomb sum for $N$ particles, $U_c=\sum_{i<j}e_ie_j/4\pi\epsilon_0
r_{ij}$, requires computation of the separations $r_{ij}$ between all
pairs of particles, which implies $\O(N^2)$ operations are needed per
sweep or time step.  Most large scale codes in use at the moment solve
Poisson's equation for the electrostatic potential using the fast
Fourier transform \cite{darden} after interpolating charges to a grid.
For an ensemble of charges interpolated to $M$ nodes of a lattice this
gives the electrostatic energy in a time which scales as $\O(M
\ln{M})$.

The use of a {\sl global} solver for the Poisson equation leads to a
strong bias in the choice of simulation techniques. Since the solution
of the Poisson equation is unique the motion of a single charge
requires the global re-calculation of the electrostatic potential, so
that Monte-Carlo methods must (apparently) lead to hopelessly
inefficient codes: {\sl All} efficient large scale codes are based at
the moment on molecular dynamics rather than Monte-Carlo.

\section{Electrostatics can be formulated locally}

It is surprising that the Coulomb interaction poses such tremendous
difficulty; after all the underlying Maxwell equations are {\sl
  local}.  The question then arises as to what part of Maxwell's
equations give rise to electrostatic interactions. Is it possible to
generate effective Coulomb interactions in a manner better adapted to
computer simulation?

In electromagnetism 
Coulomb's law comes\cite{schwinger} from a local expression for the energy:
\begin{equation}
{ U}= \int {\epsilon_0 \E^2\over 2} \dV \label{Energy}
\end{equation}
where ${\E}$ is the electric field, and the imposition of Gauss' law
\begin{equation}
\div \E - \rho/\epsilon_0 =0. 
\end{equation}
This motivates the use of the following partition function for the
electric field
\begin{equation}
 \Z(\{ \r_i\}) = \int {\cal D}\E\,\prod_\r
\delta
\left(
\div  \E - \rho\left(\{ \r_i\}\right )/\epsilon_0\right)
\,e^{- U/k_BT}.
\label{eq-Z}
\end{equation}
The charge density, $\rho({\bf r})=\sum_ie_i\delta({\bf r}-{\bf r}_i)$
and the charge of the $i$'th particle is $e_i$. We change integration
variables to 
\begin{math}  
  {\bf e} = { \bf E} +\nabla \phi_p
\end{math}
with $\nabla^2 \phi_p= -\rho(\{\r_i\})/\epsilon_0$ and find
\begin{equation}
 {\cal Z}(\{{\bf r_i}\}) = \int {\cal D}{\bf e}\; \delta(\div {\bf
  e})\, e^{ -\beta {\epsilon_0\over 2} \int ( {\bf e} - \nabla \phi_p )^2 \dV }.
\end{equation}
We now notice that the cross term in the energy is zero by integration
by parts:
\begin{equation}
 2 U/\epsilon_0 = \int ( \nabla \phi_p)^2 \dV + \int {\bf e}^2 \dV -
\underbrace {2\int \nabla \phi \cdot {\bf e} \dV }_{ {= 0}}
\end{equation}
so that
\begin{equation}
 {\cal Z}= e^{-\beta {\epsilon_0 \over 2} \int (\nabla \phi_p )^2} \int
{\cal D}{\bf e}\; \delta(\div {\bf e}) e^{ -\beta {\epsilon_0\over 2} \int {\bf
    e} ^2 \dV}
\end{equation}
We conclude that $ {\cal Z}(\{\r_i\}) = {\cal Z}_{Coulomb} (\{\r_i\})
\times {\rm const}$: Relative statistical weights are unchanged if we
allow the transverse field to vary freely, rather than quenching it to
zero.  To sample the partition function Eq.~(\ref{eq-Z}) we {\sl only}
have to find solutions to Gauss' law. If we dispense with the
condition $\curl \E=0$ usually imposed in electrostatics, $\E$ can be
{\sl any} solution to the equation $\div \E -\rho/\epsilon_0=0$; the
general solution is $\E= -\grad \phi_p + \curl \Q$ with $\Q$ arbitrary.

\subsection{Monte-Carlo sampling}

We have formulated a Monte-Carlo algorithm \cite{prl} that uses the
Metropolis method together with the energy  Eq.~(\ref{Energy}). In
order to generate configurations according to the statistical weight
of Eq.~(\ref{eq-Z}) we need to
\begin{itemize}
\item move particles without violating the constraint of Gauss' law to
  preserve the delta-function on $\div \E-\rho/\epsilon_0$;
\item integrate over the transverse degrees of freedom ${\bf e}$, of
  the electric field.
\end{itemize}

\subsubsection*{Discretization}
The system is discretized by placing charged particles on the $M=L^3$
vertices of a periodic cubic lattice, $\{i\}$. The components of the
electric field $E_{i,j}$ are associated with the $3M$ links $\{i,j\}$
of the lattice. There are $3 M$ plaquettes on the lattice each defined
by four links. We use the notation $E_{1, 2}$ to denote a local
contribution to electric flux leaving $1$ towards $2$. If the link
from $1$ to $2$ is in the positive $x$ direction we consider that this
is the local value of the $x$ component of the field $\E$. The $3M$
variables $E_{i,j}$ are thus grouped into $M$ three dimensional
vectors.

\subsubsection*{Particle motion}

\begin{figure}[htbp]
\centerline{ \psfig{file={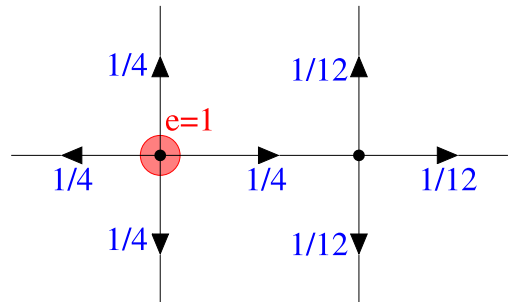},height=4cm,angle=0}}
\caption{A local solution to Gauss' law.
  Numbers on each link correspond to the electric flux in the
  direction of the arrows. There is a charge $e/\epsilon_0=1$ on the
  leftmost node of the figure.}
\label{fig:left}
\end{figure}
Two nodes of the lattice are shown in Fig.~(\ref{fig:left}). Gauss'
law is interpreted in the integral form $\int \E \cdot d{\bf S}=
e/\epsilon_0$ where $e$ is the charge enclosed by a surface enclosing
each site.  On our lattice we associate the fluxes with the links; if
a charge of $e/\epsilon_0=1$ is at the leftmost site we can satisfy
Gauss' law with a flux of $1/4$ on each outward link. On the
rightmost node the charge is zero and the signed sum of the fluxes
at this site is also zero.

\begin{figure}[htbp]
\centerline{ \psfig{file={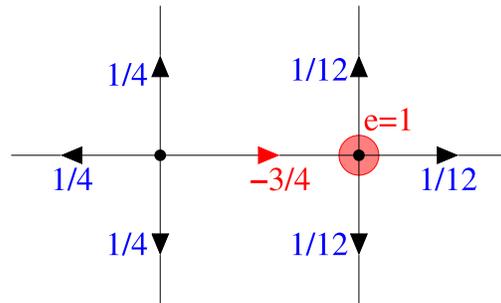},height=4cm,angle=0}}
\caption{After moving the particle from the
  leftmost to the rightmost node we wish to find a new solution to the
  Gauss constraint. This is done by modifying the flux on a single
  link connecting the two sites from $1/4$ to $-3/4$.}
\label{fig:right}
\end{figure}
We now displace the particle to the rightmost site:
Fig.~(\ref{fig:right}). We must find a new solution to Gauss' law.
{\sl One} solution is found by taking the field on the link between
the sites, $E$ and modifying it with the law $E \rightarrow E
-e/\epsilon_0$ or from $1/4$ to $-3/4$. This is the essence of the
method: Gauss' law allows an entirely local modification of the
fields, whereas Poisson's equation implies a global update of the
potential at all sites of the lattice. The energy change, $\Delta U=
\epsilon_0 \left((-3/4)^2 -(1/4)^2\right)/2$, associated with the
displacement is  used in a standard Metropolis algorithm to
decide whether to accept the update.

\subsubsection*{Transverse field }

\begin{figure}[htbp]
\centerline{ \psfig{file={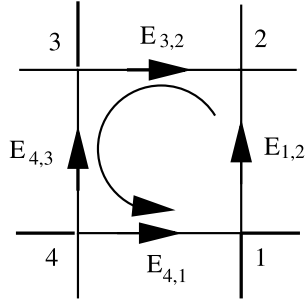},height=4cm,angle=0}}
\caption{The circulation of a plaquette is incremented by $4\Delta$ by 
  modifying the field of each link: $E_{4,3}\rightarrow
  E_{4,3}-\Delta$, $E_{4,1}\rightarrow E_{4,1}+\Delta$,
  $E_{3,2}\rightarrow E_{3,2}-\Delta$, $E_{1,2}\rightarrow
  E_{1,2}+\Delta$ }
\label{fig:plaquette}
\end{figure}

The update of the field, slaved to the particle motion is not in
itself sufficient to fully sample the partition function,
Eq.~(\ref{eq-Z}); we must also integrate over the transverse field
${\bf e}$. One can group four links into plaquettes and modify the four
links by the same increment, $\Delta$ so as to conserve Gauss' law:
Fig.~(\ref{fig:plaquette}).  In our code  particle motion and
plaquette updates are mixed stochastically with probabilities $p$ and
$(1- p)$. We chose $\Delta$ uniformly distributed in $[-\Delta_0$,$\Delta_0$], 
where $\Delta_0$ is chosen so that the Monte-Carlo
acceptance probability of this move is close to $1/2$.

\subsection{Large scale dynamics}
The above algorithm is based on local, stochastic updates for both the
particle positions and the electric field. It is thus rather natural
that both the particles and the electric field exhibit diffusive
dynamics. It can be shown\cite{dynamics} that the field dynamics are
governed by a Langevin equation:
\begin{equation}
  {\partial \E \over \partial t} = 
 D_E \left (\nabla^2 \E - \grad \chargedensity /\epsilon_0 \right)  -\J/\epsilon_0 +\curl{\zeta}(t,\r)
\label{basic}
\end{equation}
which {\sl replace} Maxwell's equations in our Monte-Carlo scheme.
The diffusion coefficient of the transverse electric field $D_E$ is
related to $p$ and $\Delta_0$.  The analytic structure of
Eq.~(\ref{basic}) is particularly interesting when there are free,
mobile charges so that the electric current is linked to the electric
field via the equation $\J=\sigma \E$. In this case the dispersion law
of the electric field develops a {\sl gap} so that
\begin{equation}
i \omega\, \hat {\bf q}\wedge \E= (\sigma/\epsilon_0 + D_E\, {\bf q}^2)\, \hat {\bf q}\wedge \E
. \label{gap}
\end{equation}

The relative diffusion rates of the electric and charge degrees of
freedom are adjusted by modifying $p$.  We initially believed that the
algorithm should be run in a regime in which the effective diffusion
coefficient of the electric field is comparable to or somewhat larger
than that of the particles: The particles are then always interacting
via a field that is close to equilibrium and are not slowed down by
the ``drag'' from the electric degrees of freedom.  In dilute systems
this choice would mean that much more CPU time must be spent updating
plaquette degrees of freedom than those corresponding to the particles
and the algorithm would become inefficient.

Simulations show that one can use a much lower rate of update for the
plaquettes.  There are several ways of understanding the efficiency of
the code \cite{auxiliary}:
\begin{itemize}
\item The gap in the electric dispersion law Eq.~(\ref{gap}) gives
  fast relaxation of the electric degrees of freedom {\sl even at long
    wavelengths} and when $D_E$ is small.
\item Motion of particles already integrates over the electric field:
  Motion of a particle around a single plaquette has {\sl exactly} the
  same effect as an update of a plaquette by $\Delta=e/\epsilon_0$.
\end{itemize}

\begin{figure}[htbp]
  \centerline{ \psfig{file={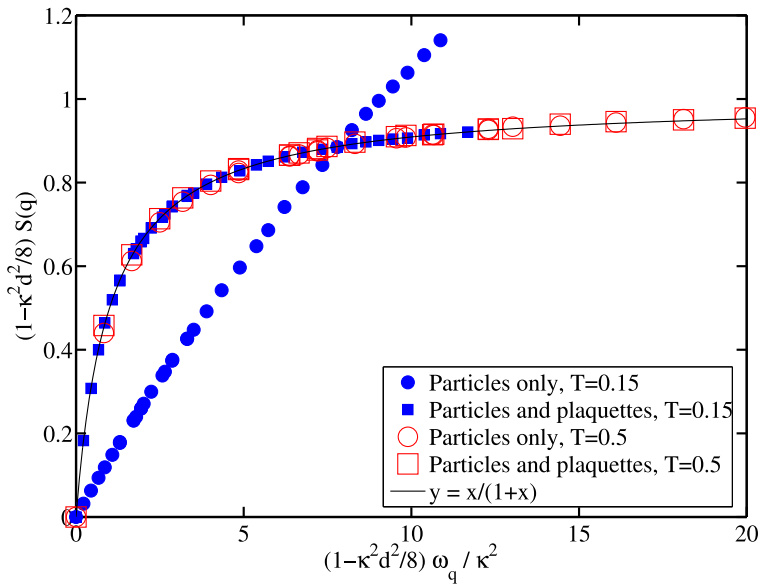},height=8cm,angle=90}}
\caption{ Charge structure factor as a function of $\omega_q$. At high temperatures
  the same structure factor Eq.~(\ref{sq}) is found with and without
  plaquette updates. At lower temperatures this law is still observed
  when the plaquettes are updated; without plaquette updates one finds
  $S\sim \omega_q$. Fit to $d=1.29$. $L=15$, $N=336$. }
\label{fig:remarkable}
\end{figure}

We simulated the above lattice gas and measured the charge-charge
structure factor.  Debye-H\" uckel theory applied to a lattice gas
gives the following expression for $S(q) = \omega_q/(\kappa^2 +
\omega_q)$ with $\omega_q=\sum_{i=1}^3 2(1-\cos{q_i})$ and $\kappa^2=
c\, e^2/k_B T\epsilon_0$, with $c$ the charge density.  When
${\bf q}\rightarrow 0$ then $\omega_q \rightarrow {\bf q}^2$.  When the finite
size of particles is also taken into account \cite{fisher} the long
wavelength structure factor must be slightly modified. We thus fit our
numerical structure factors to the expression
\begin{equation}
S(q)= {\omega_q \over \kappa^2 + \omega_q(1-\kappa^2 d^2/8)} \label{sq}
\end{equation}
where $d$ is the particle diameter.  In Fig.~(\ref{fig:remarkable}) we
show results of simulations at $k_BT \epsilon_0/e^2=0.5$ and $k_BT\epsilon_0/e^2=0.15$ which are in very
good agreement with this expression.

Numerical experimentation leads to a rather remarkable result: If one
{\sl drops the plaquette updates entirely} and works at a sufficiently
high temperature the charge-charge correlation functions are almost
indistinguishable, Fig.~(\ref{fig:remarkable}), from those in which
the system is simulated with plaquette updates.  The system is clearly
{\sl not fully ergodic}, since only values of the plaquette
circulation which are integer multiples of $e/\epsilon_0$ are
possible.  Yet charge correlation functions appear to be very similar
to those of a fully equilibrated system.

It is only at lower temperatures that there is a qualitative change in
behaviour and the system has very different behaviors depending on
whether the plaquettes are independently excited or not.  One possible
explanation is that without plaquette updates the charges are
condensing into tightly bound, neutral pairs. Indeed, if one studies a
system with just two charges at low temperatures one finds a finite,
constant line tension between charges, and thus a potential between
two charges which increases linearly with separation.

\section{Improved discretization}

The lattice gas described above is not yet useful for simulations in
condensed matter physics: The algorithm becomes {\sl very} inefficient
at low temperatures.  Motion of a particle between two nodes leads to
a finite modification of the field on the connecting link, and thus a
finite energy barrier; the Monte-Carlo acceptance rate falls off
exponentially at low temperatures. In practice this leads to
hopelessly inefficient simulations for temperatures such that $k_B
T\epsilon_0/e^2 <0.15$.  In condensed matter physics it is usual to
work in terms of a normalized Bjerrum length, $\ell_B$: the length at
which the electrostatic interaction between two particles is equal to
$k_B T$ when measured in units of particle size: $\ell_B/a= e^2/4 \pi
\epsilon_0 k_B T$. Typically we are interested in $\ell_B\sim 5-20$.
$k_B T\epsilon_0/e^2 =0.15$ corresponds to only $\ell_B\sim1/2$, we
need to gain at least a factor of ten in temperature to have a useful
code.

\subsubsection*{Charge spreading}
\begin{figure}[htbp]
\centerline{ \psfig{file={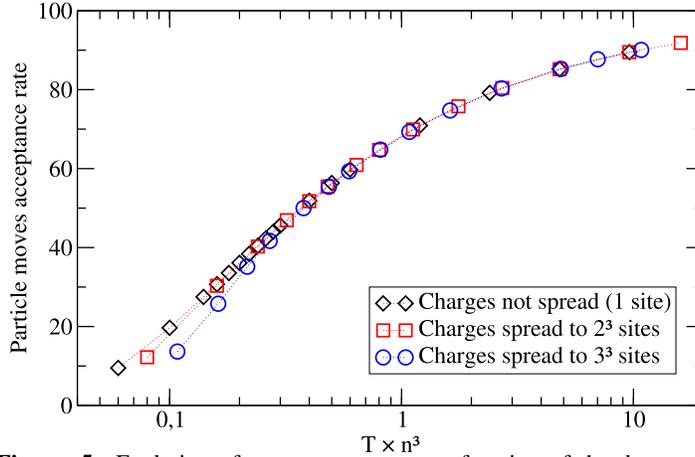},height=7cm,angle=270}}
\caption{
  Evolution of acceptance rate as a function of the degree of
  spreading. Dashed lines: guide to eye. Temperature in units of $e^2/\epsilon_0 k_B$.
}
\label{fig:accept}
\end{figure}

One can avoid the fall off in efficiency at low temperatures by
spreading the charge over several lattice sites.  In
Fig.~(\ref{fig:accept}) we have plotted the Monte-Carlo acceptance
rate for particle motion as a function of temperature for a system of
dimensions $L=15$ with two free charges. We associate the charge of
each particle to $n^3$ sites with $n=1,2,3$. We find that the
temperature at which the code becomes inefficient varies as $T\sim
1/n^3$. When we measure the Bjerrum length in terms of the size of the
spread-out particle we find that we can simulate at a Bjerrum length
$\ell_B < n^2/2$ without the acceptance rate falling too low.

\subsubsection*{Off-lattice methods}

By allowing the particles to move in the continuum and then
interpolating the charges to the network (as is usual in Fourier codes
\cite{darden}) we can continuously tune the Monte-Carlo step size so
that the acceptance rate does not fall off with temperature.  Many
choices of interpolation scheme are possible, in our codes
\cite{rottler1} we have used a scheme based on piecewise quadratic
splines, leading to interpolation to a cube of $27$ sites around each
particle.

\begin{figure}[htbp]
\centerline{ \psfig{file={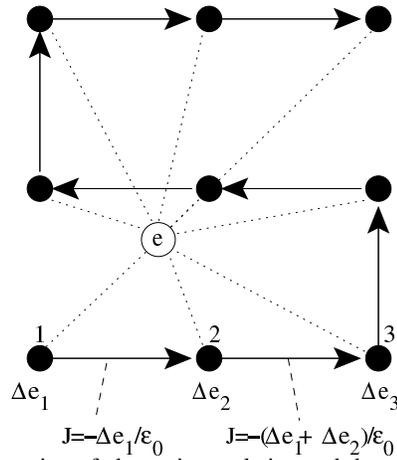},height=6cm,angle=0}}
\caption{Illustration of charge interpolation and the
  current update using a Hamiltonian path in two dimensions. A charge
  (open circle) interpolates onto the $3^2=9$ lattice sites (solid).
  The partial charges on these sites change by an amount $\Delta e_i$
  when the particle moves. The current $\J$ flows along the Hamiltonian
  path and terminates at the last site, since $\sum_i\Delta e_i=0$.  }
\label{fig:hamilton}
\end{figure}

When a particle moves in the continuum we need to generalize the
update rule $E\rightarrow E-e/\epsilon_0$ that generates the local
updates to the electric field. We must generate an electric current
$\J$ which is a solution to the continuity equation $\div \J= -\Delta
\rho$, where $\Delta \rho$ is the finite variation of the charges on
the nodes calculated from the spline 
interpolation scheme.  We then update the electric field with $\E
\rightarrow \E -\J/\epsilon_0$. The equation for $\J$ is clearly {\sl
  under-determined}. We use this fact to our advantage to find {\sl
  one} solution which is simple.  This is illustrated in
Fig.~(\ref{fig:hamilton}). We construct a Hamiltonian path which
visits each node where the interpolated charge has changed. The
current is then calculated on each link as the sum of the charge
variations on all previous nodes on the path.

%Going off-lattice does slow the algorithm down: the calculation of the
%splines is a major overhead compared with the simple lattice gas.
%However, for the large simulations which are now currently performed
%with $N>10^5$ it is still faster than direct evaluation of the pair
%potential.

\section{Cluster sampling for the electric field}

In many physical situations the algorithm as described above is
remarkably efficient: The presence of a finite concentration of free
charges always gives rise to a finite $\sigma$ and thus fast field
relaxation: the slowest modes in the system are then modes of density
relaxation. However, one is sometimes interested in highly
heterogeneous systems where charges are either not mobile (perhaps
they are attached to a polymer or a surface) or are excluded from
certain parts of the simulation volume (perhaps due to the presence of
a large colloidal inclusion).  Similarly in dipolar fluids such as
water in the absence of free charges $\sigma=0$.  In this case we have
to rely on the slower diffusive motion of the electric field in
Eq.~(\ref{basic}) driven by the plaquette updates. We will now show
how to generate updates which give rise to fast field equilibration
even for the case $\sigma=0$ using a cluster algorithm for the
electric field variables.

There is a remarkable analogy between the constrained partition
function Eq.~(\ref{eq-Z}) and the bosonic 2+1 dimensional Hubbard
model at zero temperature in the phase approximation \cite{dual}.
Its partition function is written in terms of a constrained vector
field discretized on a lattice:
\begin{equation}
\sum_{ \E\, {\rm integer} } \exp{\left( - {1\over 2K} \int \E^2 \dV  \right)}\;
\prod_{\r} \delta (\div \E). \label{hubbard}
\end{equation}
The major difference compared with the partition function
Eq.~(\ref{eq-Z}) is that the field variables on each link are now
integers rather than reals; the field variables represent the real
bosonic currents in the system.  $K$ is the ratio between kinetic
energy and the repulsive energy and is interpreted as an effective
temperature. The partition function Eq.~(\ref{hubbard}) has been
extensively studied \cite{dual} and has a phase diagram with a small
$K$ (large repulsion) insulating phase and a superfluid phase at large
$K$ (large kinetic energy).There is a phase transition between these
phases at $K=0.33305(5)$.  We have already seen an example of a
similar partition function when we suppressed the plaquette updates in
the electrostatic algorithm. The very different structure factors
observed in Fig.(\ref{fig:remarkable}) are presumably linked to the
two phases of the Hubbard model.

Early numerical studies of Eq.~(\ref{hubbard}) were based on local
updates to plaquette degrees of freedom, in a manner very similar to
that described above for the electrostatic algorithm. More recently
%JR cut IT
{\sl much} more efficient algorithms have been found based on cluster
or worm sampling \cite{fabien}.  The algorithm works by nucleating a
pair of positive and negative pseudo-particles on the same site of the
system. One of the pseudo-particle is then moved on the lattice so
that $\div \E \ne 0$. The motion of this pseudo-particle is a random
walk biased by the energy in Eq.~(\ref{hubbard}) 
%in such a way that it
%performs a {\it non-local} move with {\it local} stochastic decisions
\cite{fabien}.  The pseudo-particle eventually returns to
its stationary partner and the two particles then annihilate each
other.  When the pseudo-particle moves, the electric field is updated
exactly as described above for the electrostatic algorithm,
$E\rightarrow E- e/\epsilon_0$.  At the moment of annihilation the
total weight of the generated path or cluster is compared with its
time reversed version and the update is either globally accepted or
refused \cite{fabien}.  The algorithm is very reminiscent of the
creation of virtual pairs of particles in quantum mechanics.  In the
original version of the algorithm \cite{fabien} the pair of
pseudo-particles had a charge $e/\epsilon_0=\pm 1$ in order to
generate fields with integer flux on the links.

\begin{figure}[htbp]
\centerline{ \psfig{file={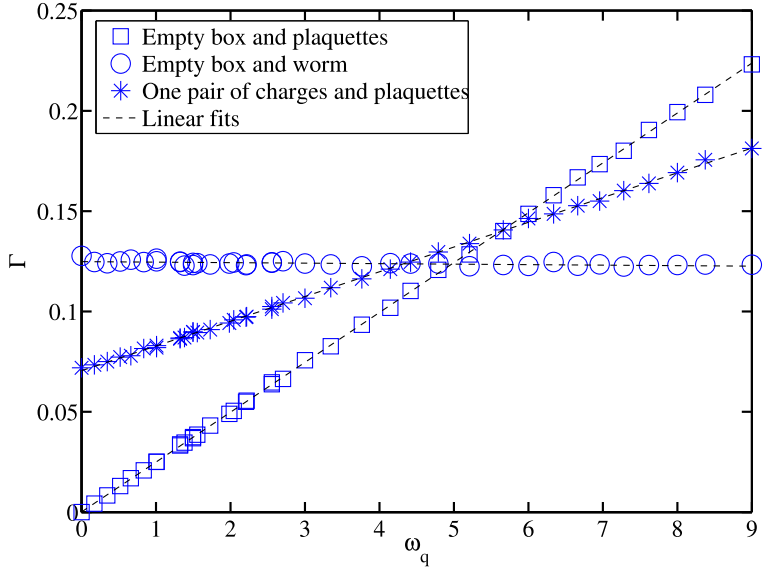},height=8cm,angle=90}}
\vskip .5cm
\caption{
  Decay rate, $\Gamma$ of the transverse electric field as a function
  of $\omega_q$. $\Box$: empty box, plaquette updates only
  illustrating diffusive field motion $\Gamma= D_E\omega_q$; time
  scale: $L^3$ Monte-Carlo tries.  \Pisymbol{pzd}{83}: two particles in box with
  plaquette updates $\Gamma= \sigma/\epsilon + D_E \omega_q$; time scale:
  $L^3$ Monte-Carlo trials, 1:1 split in Monte-Carlo attempts between
  particles and plaquettes.  $\ocircle$: cluster algorithm showing
  uniform relaxation across all modes; time scale: a single cluster
  update.  $L=15$, $k_BT\epsilon_0/e^2=0.5$ . }
\label{fig:rates}
\end{figure}

Given the efficiency of this algorithm in sampling Eq.~(\ref{hubbard})
we have explored the interest of the method for sampling the
electrostatic partition function Eq.~(\ref{eq-Z}).  In the
electrostatic problem we are interested in integrating over all values
of the electric field. Thus we create the pair of virtual particles
with a random charge distributed according to a uniform distribution
between $-e_m$ and $e_{m}$. If $e_{m}$ is small we find that the
typical cluster is very large containing about $0.7 \times L^3$ links.
In order to maximize the rate of change in the field on the links we
should use a large value for $e_{m}$. However, the use of a too large
value for $e_{m}$ prevents the propagation of the virtual particle,
which backtracks too readily in its own path and only visits $\O(1)$
sites before the annihilation.  This is again understood in the boson
analogy as being due to the different behaviour of a particle in a
superfluid or insulating phase.  In practice we use $e_{m}^2 \sim k_BT
\epsilon_0$.

In Fig.~(\ref{fig:rates}) we plot the decay rate for the Fourier
components of the transverse electric field for three different
simulations. The first simulation is a box without free charges
simulated using simple plaquette updates. As predicted by
Eq.~(\ref{basic}) the dynamics of the field is diffusive, $\Gamma =
D_E q^2$ for small $q$.  We then add two free charges to the system
and measure the same relaxation rate.  We see that a finite gap is
introduced into the dispersion law as predicted by Eq.~(\ref{gap}).
Finally we simulate the system without charges using the cluster
algorithm.  We find that the relaxation rate is weakly dependent on
the mode and we have a method which equilibrates the transverse
electric field in about two cluster moves per link, rather than the
$\O(L^2)$ sweeps needed for the diffusion of the electric field with
local plaquette updates.  Mixing particle motion with occasional
cluster updates in a heterogeneous system now allows one to ensure
fast relaxation of the electric degrees of freedom even in situations
where $\sigma$ is zero.

We understand the efficiency of the cluster algorithm in both the
Hubbard model and in the electrostatic algorithm as being due to the
dispersion law Eq.~(\ref{basic}) in the presence of a current
$\J=\sigma \E$. We have seen that if $\sigma$ is non-zero then we
develop a gap in the electric spectrum and the system is described by
a dynamic exponent $z=0$ rather than $z=2$ characteristic of
diffusion. In the cluster algorithm the finite conductivity comes from
the virtual particles rather than from true free ions but their effect
is very similar on the relaxation spectrum for the electric field.

\end{document}